\documentclass[aps,prl,twocolumn,preprintnumbers,10pt,showpacs]{revtex4-1}
\usepackage{amsmath,bm,amssymb}

\usepackage[utf8]{inputenc}

\usepackage[hyperindex,breaklinks,hyperfootnotes=false,bookmarks=false]%
{hyperref}

\usepackage{graphicx}
\usepackage{subcaption}
\usepackage{psfrag}
\usepackage{xcolor}
\usepackage{soul} 
\usepackage{bbold}

\renewcommand{\tabcolsep}{2em}

\DeclareRobustCommand{\fig}[1]{Fig.~\ref{fig:#1}}
\DeclareRobustCommand{\eq}[1]{eq.~\eqref{eq:#1}}

\newcommand\Cfour{\frac{d_R d_X}{N_R}}
\newcommand\CfourF{\frac{d_R d_F}{N_R}}
\newcommand\CfourA{\frac{d_A d_A}{N_A}}

\newcommand{\ord}[1]{\mathcal{O}\!\left(#1\right)}

\newcommand\eps{\epsilon}

\def \tr {\mathop{\rm tr}\nolimits}

\usepackage{slashed}

\allowdisplaybreaks

\begin{document}

\preprint{MPP-2019-3, ZU-TH 01/19, MITP/19-001} 

\title{Matter dependence of the four-loop cusp anomalous dimension}

\author{J.\ M.\ Henn$^{a}$, T.\ Peraro$^{b}$, M.\ Stahlhofen$^{c}$, P.\ Wasser$^{c}$}

\affiliation{
$^a$ Max-Planck-Institut für Physik, Werner-Heisenberg-Institut, 80805 München, Germany\\
$^b$ Physik-Institut, Universit{\"a}t Zürich, Wintherturerstrasse 190, CH-8057 Zürich, Switzerland\\
$^c$ PRISMA Cluster of Excellence, Institute of Physics, Johannes Gutenberg University, D-55099 Mainz, Germany
}
\pacs{12.38Bx}

\begin{abstract}
We compute analytically the matter-dependent contributions to the
quartic Casimir term of the four-loop light-like cusp anomalous
dimension in QCD, with $n_f$ fermion and $n_s$ scalar
flavours.  The result is extracted from the double pole of a scalar form
factor.  We adopt a new strategy for the choice of master integrals 
with simple analytic and infrared properties, which significantly simplifies our calculation.
To this end we first identify a set of integrals whose integrands have a dlog form,
and are hence expected to have uniform transcendental weight.  
We then perform a systematic analysis of the soft and
collinear regions of loop integration and build linear combinations of
integrals with a simpler infrared pole structure. In this way, only integrals
with ten or fewer propagators are needed for obtaining the cusp
anomalous dimension. These integrals are then computed via the method
of differential equations through the addition of an auxiliary
scale. Combining our result with that of a parallel paper, we obtain
the complete $n_{f}$ dependence of the four-loop cusp anomalous
dimension in QCD. Finally, using known numerical results for the gluonic contributions,
we obtain an improved numerical prediction 
for the cusp anomalous dimension in $\mathcal{N}=4$ super Yang-Mills theory.
\end{abstract}

\maketitle

\section{Introduction}

The cusp anomalous dimension is a universal quantity appearing in QCD.
It governs infrared divergences of scattering amplitudes \cite{Mueller:1981sg,Korchemsky:1991zp,Korchemskaya:1994qp}, and appears in the large
spin limit of twist-two operators \cite{Korchemsky:1992xv}. It also controls the resummation of large Sudakov double logarithms due 
to soft and collinear emissions and is therefore relevant to many collider observables, see e.g. \cite{Sterman:1986aj,Becher:2006mr,Becher:2008cf,Abbate:2010xh,Hoang:2015hka,Stewart:2013faa,Becher:2013xia,Chen:2018pzu}.
Its colour dependence is governed by non-Abelian exponentiation, which
allows for the first time for quartic Casimir terms at four loops. The latter
have received a lot of attention, in particular because their presence implies
a breaking of the Casimir scaling property, and since they represent the last missing
four-loop ingredient in the above calculations.
Further interest comes from the fact that these are the first truly non-planar
terms in $\mathcal{N}=4$ super Yang-Mills (sYM). In this theory, the planar cusp anomalous
dimension is known from integrability \cite{Beisert:2006ez}, and it remains an open question whether 
integrability extends to the non-planar sector.

The planar cusp anomalous dimension is known to four loops in QCD \cite{Henn:2016men}, and some of the $n_{f}$ dependent 
terms have been computed \cite{Lee:2016ixa,Davies:2016jie,vonManteuffel:2016xki,Lee:2017mip,Moch:2017uml}. 
Recently, numerical results were obtained for the quartic Casimir terms in $\mathcal{N}=4$ sYM \cite{Boels:2017ftb,Boels:2017skl}
and in QCD \cite{Moch:2018wjh}. In this Letter, we present the first analytic result for the $n_{f}$ terms in QCD.

Given the complexity of such a non-planar four-loop calculation, we develop and use cutting-edge methods
to achieve this goal. The latter may be of interest in their own right, as we expect they can be applied to many
other situations.

We use as our starting point a form factor of composite operators inserted into two on-shell states.
Thanks to the universality of the cusp anomalous dimension, we are free to choose a suitable operator,
and we make a particularly simple choice, as explained below.
The kinematic dependence is fixed by dimensional analysis, so that the form factor essentially
depends on $\eps$, the parameter of dimensional regularization in $D=4-2\eps$ dimensions, only.

In recent years, it has become standard to make an educated choice of Feynman integral basis \cite{ArkaniHamed:2010gh,Henn:2013pwa,WasserMSc}, 
where the integrals are of uniform transcendental weight (UT), or so-called pure functions. A given $L$-loop Feynman integral with this property
has the $\eps$ expansion $I_{\rm pure} = \eps^{-2 L} \sum_{k} c_{k} \eps^k$, where the $c_{k}$ are numbers of transcendental weight $k$.
This property is particularly useful in $\mathcal{N}=4$ sYM where, conjecturally, the form factors have uniform and maximal weight.
In general the form factors are expressed as $ F = \sum_{i} r_{i}(\eps) I_{i \; \rm pure}$ with some rational functions $r_{i}(\eps)$, however, in the latter theory the $r_{i}$ are just numbers, i.e. $\eps$-independent. 
Note that this property only becomes visible when a basis of pure functions is chosen. 
One of the first applications of these ideas was at the level of the three-loop form factor \cite{Gehrmann:2011xn}.
We argue that such a basis choice will be of crucial importance also in QCD. Experience from lower loops shows that terms having at least one factor of $n_{f}$ have a drop of transcendental weight. In a UT basis, this means that all coefficients have the form $r_{i}(\eps) = \eps q_{i}(\eps)$.
Making this property manifest allows us to take a calculational shortcut.

The quartic Casimir terms appear for the first time at four loops and, as a consequence of renormalizability, they come with a $1/\eps^2$ pole (whose coefficient is the cusp anomalous dimension). Thanks to the additional factor of $\eps$ mentioned above, we need to know the four-loop integrals only up to (and including) the $1/\eps^3$ pole. 
In order to take advantage of this fact, we classify the pure functions according to their soft and collinear divergence properties \cite{ArkaniHamed:2010gh,Drummond:2010mb,Bourjaily:2011hi}.
In this way, we can arrange integrals having many propagators into linear combinations that have only $1/\eps^2$ or better pole structure, and hence are irrelevant for the determination of the cusp anomalous dimension. In this way, only a subset of form factor integrals is needed.

Expressions for all planar four-loop form factor integrals were obtained previously \cite{Henn:2016men} by an application of the differential equations method \cite{Kotikov:1990kg,Bern:1993kr,Gehrmann:1999as,Henn:2013pwa,Henn:2013nsa}.
Here, we evaluate all required non-planar integrals using the same method.

\section{Setup and definitions}

We work in massless QCD with gauge group $SU(N_{c})$ and $n_{f}$ fermion flavors. 
For convenience, we couple the theory canonically, i.e. through covariant derivatives, to $n_{s}$ complex scalar fields,
with canonical kinetic term $\phi \square \bar{\phi}$. 
This allows us to consider a composite
operator $\mathcal{O} = \phi  \bar\phi$, inserted into on-shell scalar states, i.e. with $p_1^2 = p_2^2=0$,
\begin{align}
  F=   \, \langle \mathcal{O} \,  {\phi}(p_1)  \bar{\phi}(p_2)\rangle
 \,.
 \label{eq:FFdef}
 \end{align}
 Here the scalar fields are considered to be in the representation $R$ of $SU(N_{c})$,
 which we take either to be the fundamental (F), or adjoint (A).
 In the following, we will set the only kinematic scale $ 2 p_1 \cdot  p_2  = -1$, and the dimensional
 regularization scale $\mu^2=1$, without loss of generality.
The fact that $\mathcal{O}$ has spin zero
means that in momentum space, no additional momentum operator is inserted into 
the diagram at the cusp. As a consequence, 
the corresponding Feynman diagrams contain one numerator factor less compared to what one would have obtained e.g. for a fermion current $\mathcal{O}^{\mu} = \bar{\psi} \gamma^{\mu} \psi$. 

The cusp anomalous dimension is universal, that means it does not depend on the 
types of external particles, in this
case scalars. 

\begin{figure*}[t]
\begin{center}
\includegraphics[width=0.29 \textwidth]{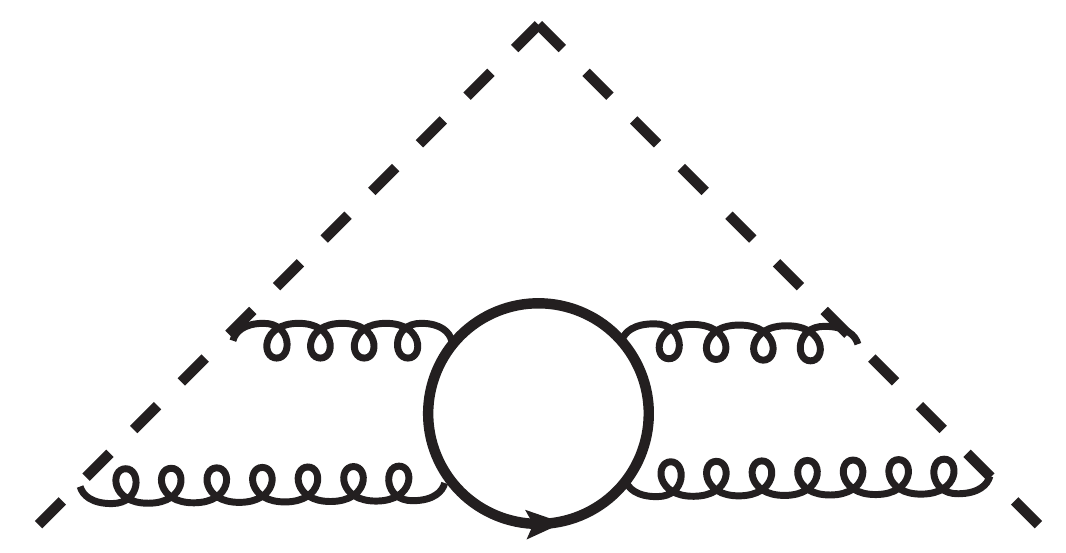}%
\put(-130,55){(A)} \;
\includegraphics[width=0.29 \textwidth]{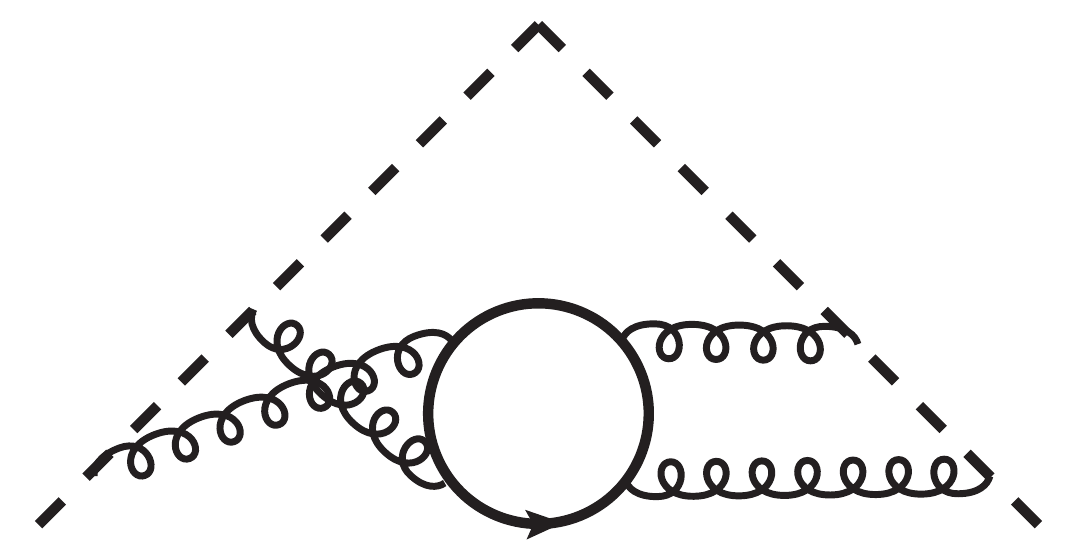}%
\put(-130,55){(B)} \;
\includegraphics[width=0.29 \textwidth]{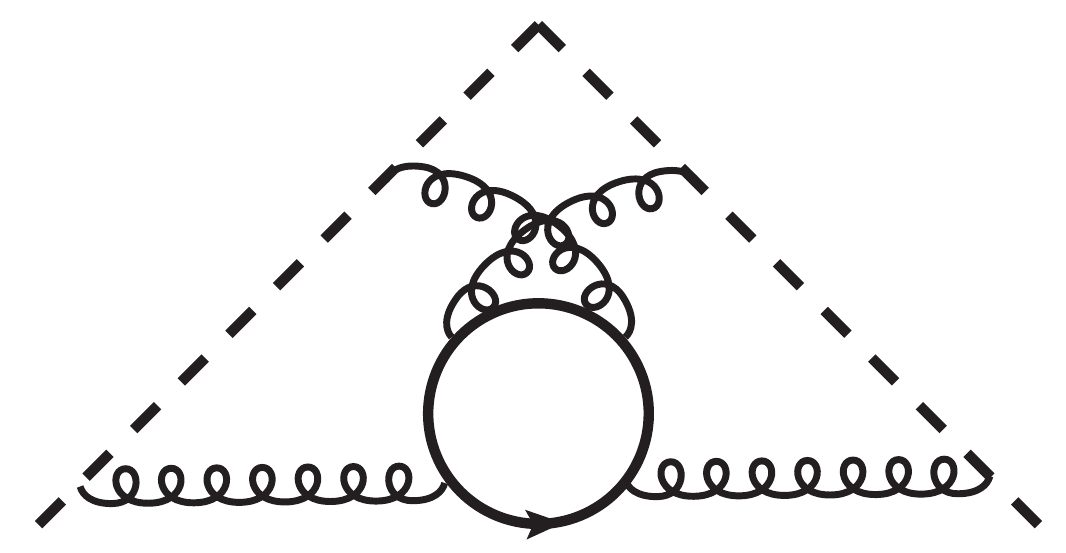}%
\put(-130,55){(C)}
\\
\includegraphics[width=0.29 \textwidth]{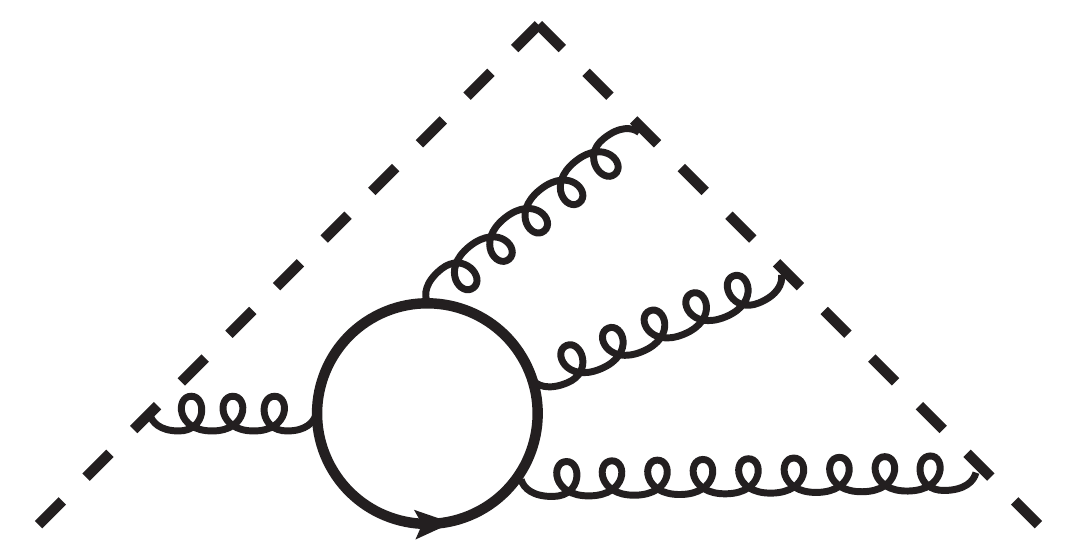}%
\put(-130,55){(D)} \;
\includegraphics[width=0.29 \textwidth]{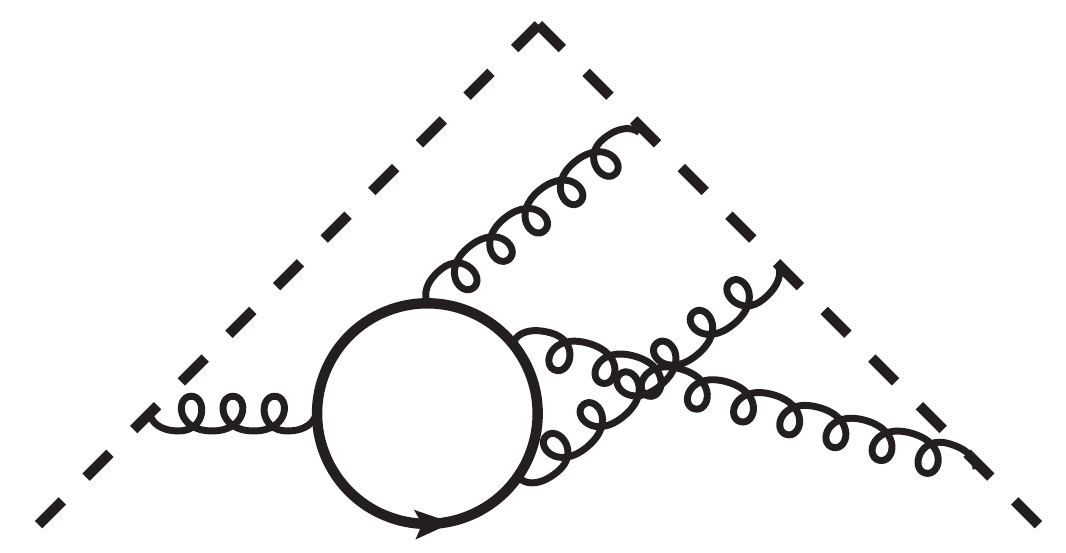}%
\put(-130,55){(E)} \;
\includegraphics[width=0.29 \textwidth]{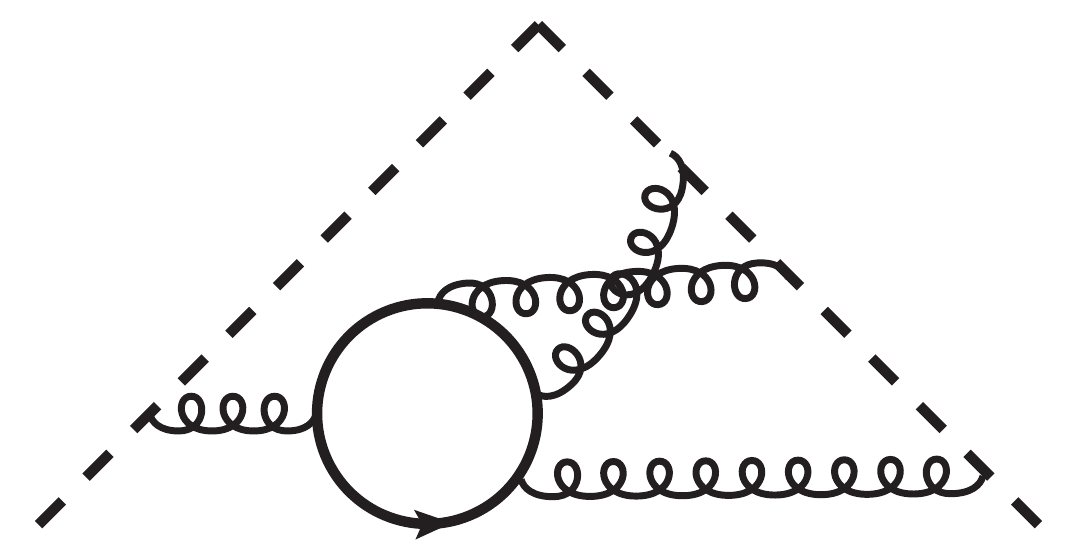}%
\put(-130,55){(F)}
\end{center}
\caption{Feynman diagrams contributing to the $n_f {d_R^{abcd} d_F^{abcd}}/{N_R}$ term of the four-loop cusp anomalous dimension.
These diagrams also define the top sector topologies of the associated integral families listed in Table \ref{tab:integrals}.
\label{fig:diags}
}
\end{figure*}

We are interested in the four-loop contribution to $F$ with the quartic Casimir structure \cite{vanRitbergen:1997va}
\begin{align}
\Cfour \equiv \frac{d_R^{abcd} d_X^{abcd}}{N_R}\,,
\end{align}
where 
$d_R^{abcd} = \tr_R\bigl[T_R^{(a} T_R^{b\vphantom{(}} T_R^{c\vphantom{(}} T_R^{d)}\bigr]$, $n_R=\tr_R \mathbb{1}$, and $X$ denotes the $SU(N_c)$  representation of the internal matter fields ($n_f$ fermions and $n_s$ scalars).
The quartic Casimir $n_f$ and $n_s$ contributions originate from a small set of four-loop Feynman diagrams with an internal fermion box, as shown in Fig.~\ref{fig:diags}, and internal scalar box, triangle, and bubble subdiagrams, respectively.
There is also a corresponding gluonic quartic Casimir term, which however is beyond the scope of the present paper.

The general structure of infrared divergences of the form factor,
together with the fact that the quartic terms appear for the first time at this loop order, implies that 
\begin{align}
F|_\mathrm{quartic}^{n_f,\,n_s} 
&=
-\frac{1}{32}  \,\frac{1}{\eps^2} \, \bigg(\frac{\alpha_s}{\pi}\bigg)^4 K_4|_\mathrm{quartic}^{n_f,\,n_s}
+\ord{\eps^{-1}}\,,
\label{eq:FFcusp}
\end{align}
where $\alpha_s$ is the strong coupling.
Our goal is to determine $K_4|_\mathrm{quartic}^{n_f,\,n_s}$.
We perform the calculation in a general covariant gauge with parameter $\xi$, 
and we verify that the linear terms in $\xi$ disappear from the result.
\section{Integral reduction}
The form factor $F|_\mathrm{quartic}^{n_f,\,n_s}$  is expressed in terms of scalar Feynman 
integrals. 
A first step in its calculation consists in exploiting integration-by-parts identities (IBP) \cite{Chetyrkin:1981qh} in order to reduce the expression 
to a minimal number of so-called master integrals (MI). 
We find that even on state of the art compute servers, publicly available IBP reduction programs run into difficulties.

For this reason, we use novel techniques pioneered
in~\cite{vonManteuffel:2014ixa,Peraro:2016wsq}.  We generate the system of IBP equations and
mappings for each topology, using
\textsc{LiteRed}~\cite{Lee:2012cn}.  
We then solve it modulo prime numbers using a custom linear solver for sparse systems over finite fields, 
and reconstruct the full analytic result using the techniques illustrated in~\cite{Peraro:2016wsq}. 
Equivalences between integrals appearing in different integral families are identified using {\tt TopoID}~\cite{Hoff:2016pot}.

\section{Master integrals and pure functions}
\begin{table}
\renewcommand{\tabcolsep}{6pt}
\begin{tabular}{c|cccc|c|c|c}
family& $\le 9$ & 10 & 11 & 12 & $\Sigma$ & $\Sigma^*$ & dlogs\\
\hline
A & 39 & 5 & 0 & 1 & 45 & 45 & 170\\
D & 33 & 1 & 1 & 0 & 35 & 5 & 66 \\
B & 38 & 5 & 0 & 2 & 45 & 21 & 194\\
C & 53 & 16 & 0 & 2 & 71 & 21 &  305\\
E & 32 & 2 & 1 & 0 & 35 & 5 & 88 \\
F  & 38 & 2 & 1 & 0 & 41 & 3 & 94
\end{tabular}
\caption{Master integrals (MI) by integral family, total number of MI, and number of {\it dlog} integrands found.}
\label{tab:integrals}
\end{table}
Table \ref{tab:integrals} gives an overview of the number of MI for each of the integral families.
The second to fourth columns state the number of MI per family, grouped according to the number of propagators.
$\Sigma$ is the total number of MI of the corresponding family, whereas $\Sigma^*$ is the number of 
MI excluding all integrals that can be related to integrals of an integral family previously considered (i.e., that appears above in the same table).
We ordered the families such that the first two families are the planar topologies and then we have the non-planar topologies.
In total, adding all entries for $\Sigma$, we have 272 MI. After considering all relations between 
MI this number is reduced to 100 (the sum of all entries in the $\Sigma^*$ column).

It is advantageous to select a basis of pure integrals.
Conjecturally, the latter can be identified by checking that their four-dimensional loop {\it integrands}
can be put into a so-called {\it dlog} form \cite{Arkani-Hamed:2014via}.
(See also \cite{Chicherin:2018old} for recent developments on the topic of identifying UT integrals.)
We systematically find such integrals using the algorithm \cite{WasserMSc}.
The last column of Table \ref{tab:integrals} gives the number of {\it dlog} integrals we found in the different families,
based on an ansatz with heuristic power counting constraints. 
We find that it is possible to choose a subset of these {\it dlog} integrals 
that can be used as a complete basis of MI.

Using this basis (denoted by $f$) to express the result, we find 
\begin{align}\label{eq:Fweightdrop}
F|_{\rm quartic}^{n_f,\,n_s}  =\Cfour C(n_f,n_s) \, \sum_{i=1}^{100} \eps q_{i}(\eps) f_{i}\,,
\end{align}
where $C(n_f,n_s)$ denotes the overall normalization for the case of internal fermions or scalars and the $q_i$ are IBP coefficients of $\mathcal{O}(\eps^0)$.
Remarkably, all integral coefficients are proportional to $\eps$!
This confirms our expectation that $F|_\mathrm{quartic}^{n_f,n_s}$ has a transcendental weight drop.
Note that it is essential to use a basis of pure functions to observe this property
prior to computing the integrals.

\section{Integrals with better IR properties}

The pure functions we found in the previous section may have several (in general, nested)
regions of soft and collinear divergence, due to the on-shell light-like kinematics.
At four loops, these regions lead to poles of up to $\eps^{-8}$. On the other hand, we expect the
quartic Casimir contribution to be given by a $\eps^{-2}$ pole only, see eq. (\ref{eq:FFcusp}).

This motivates the question of whether the infrared structure of the four-loop {\it integrand} can
be made manifest. In \cite{ArkaniHamed:2010gh,Drummond:2010mb,Bourjaily:2011hi,Eden:2012tu}, integrands for scattering amplitudes and correlation functions (see also \cite{Anastasiou:2018rib})
were constructed such that certain one-loop soft and collinear regions (and hence the associated divergences) are suppressed.
Here, we perform a dedicated, algorithmic analysis of all $L$-loop soft or collinear regions of the four-loop integrands. This information allows us to construct loop integrals for which we can give an upper bound
on the degree of divergence. 

In the following we briefly sketch the implementation of this algorithm.
In order to test the region, where the loop momentum $k_i$ becomes collinear to 
$p_1$ ($\gamma_{1,i}\!\to\! 0$) and/or soft ($\beta_i\!\to\! 0$), we 
parameterize
\begin{align}
 k_i^\mu &= \beta_i \, p_1^\mu + \beta_i \, \gamma_{1,i}^2 \, p_2^\mu + \beta_i 
\,
\gamma_{1,i} \,\bar{k}_{\perp i}^\mu \,, 
\label{eq:p1csoftparam}
\end{align}
with $p_1 \cdot \bar{k}_{\perp i} = p_2 \cdot \bar{k}_{\perp i}=0$, and 
analogously for $k||p_2$.
We also consider consecutive $p_1$- and $p_2$-collinear limits of $k_i$ 
($\gamma_{1,i} \!\to\! 0$ and $\gamma_{2,i}\!\to\! 0$, respectively) using the 
parametrization 
\begin{align}
 k_i^\mu &= \gamma_{2,i}^2 \, p_1^\mu + \gamma_{1,i}^2 \, p_2^\mu + 
\gamma_{1,i} 
\,
\gamma_{2,i} \,\bar{k}_{\perp i}^\mu 
\,.
\label{eq:p1cp2cparam}
\end{align}

We can now take soft and collinear limits of each loop momentum separately in arbitrary order.
We do this by Laurent expanding the integrand in the soft and collinear parameters $\beta_i$ and 
$\gamma_{1/2i}$, respectively.
If we find a single pole of the form $d \beta_i/\beta_i$ or $d 
\gamma_{1/2i}/\gamma_{1/2i}$ we conclude that the corresponding limit 
(potentially) contributes a $1/\eps$ pole to the integral.
We then proceed with the residue of this pole and test the next limit, and so on.
Note that the {\it dlog} property guarantees that we never encounter more than
single poles in this procedure.

\begin{figure}[t]
\includegraphics[width=0.25 \textwidth]{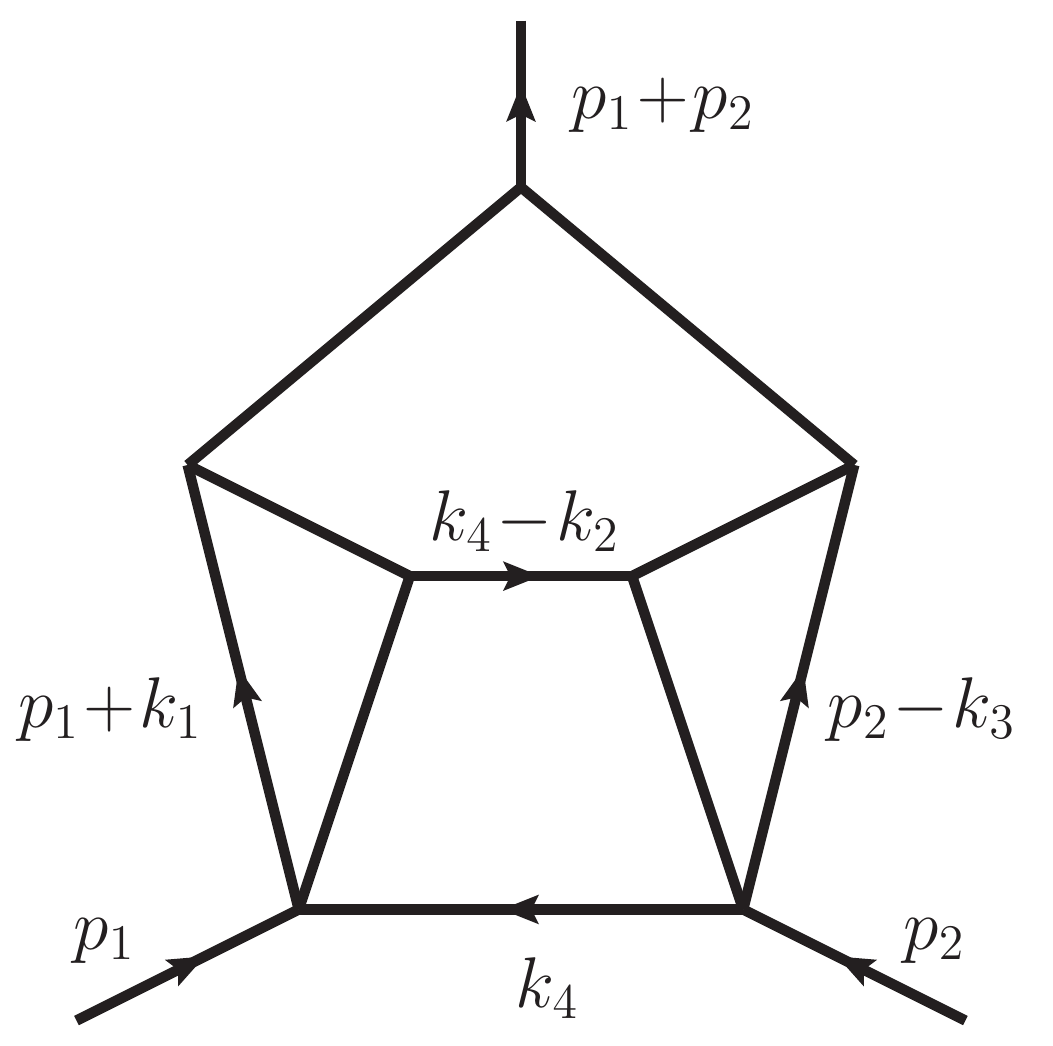}%
\caption{Integral used to illustrate our infrared analysis.
\label{fig:example}}
\end{figure}

Our code systematically checks all consecutive soft or collinear 
limits of the $k_i$.
As an example, consider the  Feynman integral shown in \fig{example}.
According to our algorithm, the maximal singular behaviour comes from first 
taking the joint $p_1$-collinear limit of loop momenta $\{1, 2, 4\}$ 
($\gamma_{1,124} \to 0$), then $\gamma_{1,3} \to 0$ followed 
by the joint limit $\gamma_{2,234} \to 0$ and finally $\gamma_{2,1} \to 0$.
Hence, we expect at most a fourth pole in $\eps$.
This is indeed confirmed by the available analytic result \cite{Henn:2016men}:
$\pi^4/(5184 \eps^4) + \mathcal{O}(\eps^{-3})$.

In the case of planar integrands, it is sufficient to perform the above analysis 
for one (canonical) momentum routing, corresponding to region coordinates.
For nonplanar integrals we adopt a pragmatic approach and run the algorithm for all momentum routings 
where 4 of the 12 propagators of the nonplanar topologies coincide with $1/k_i^2$.

Our method gives an upper bound on the degree of divergence of an integral only,
as there may be cancellations, or some regions may give zero contributions, e.g. due to scale-less
integrals. Note that we make the physical assumption that only soft and collinear regions are relevant to this analysis, so that the potential presence of other scaling regions could alter the conclusions.
For all integrals that are known or that we explicitly computed, we verified that our bound was satisfied, thereby validating the procedure. 

We use the information on the infrared behaviour of the individual integrals to
assemble a {\it dlog} basis where integrals having more than 10 propagators have at most $1/\eps^2$ poles.
As all {\it dlog} integrals appear in $F$ with a coefficient that is $\mathcal{O}(\eps)$, cf. eq. (\ref{eq:Fweightdrop}), this implies that we do not need
integrals with more than 10 propagators  in order to extract $K_4|_\mathrm{quartic}^{n_f,\,n_s}$.

\section{Computation of master integrals}

We use the method of computing Feynman integrals via differential equations in canonical form \cite{Henn:2013pwa}, adapted to form factor integrals in \cite{Henn:2013nsa} (and used in subsequent work \cite{Lee:2016ixa,Lee:2017mip}). The idea is to introduce a second scale, e.g. $p_{1}^2 \neq 0$, so that the the form factors have a non-trivial kinematic dependence, which is computed via differential equations. Note that as $p_{2} \to 0$, the integrals degenerate to propagator-type integrals that are all known. Therefore one can use the differential equations to determine the desired on-shell form factor integrals by relating them to the known propagator-type integrals \cite{Lee:2011jt}. 

In this way we obtain all required integrals analytically, up to transcendental weight 8.
The necessary integral reductions are performed using \cite{Smirnov:2008iw,Peraro:2016wsq}.
We remark that it follows from the form of the differential equations that the equations relating the results of the propagator-type integrals to our form factor integrals involve only harmonic polylogarithms \cite{Remiddi:1999ew,Maitre:2005uu}
with indices $0$ and $1$, evaluated at $1$. As a result, only multiple zeta values appear in these equations, to any order in $\eps$. 
The results are provided in ancillary files.

 \begin{figure}[t]
\includegraphics[width=0.25 \textwidth]{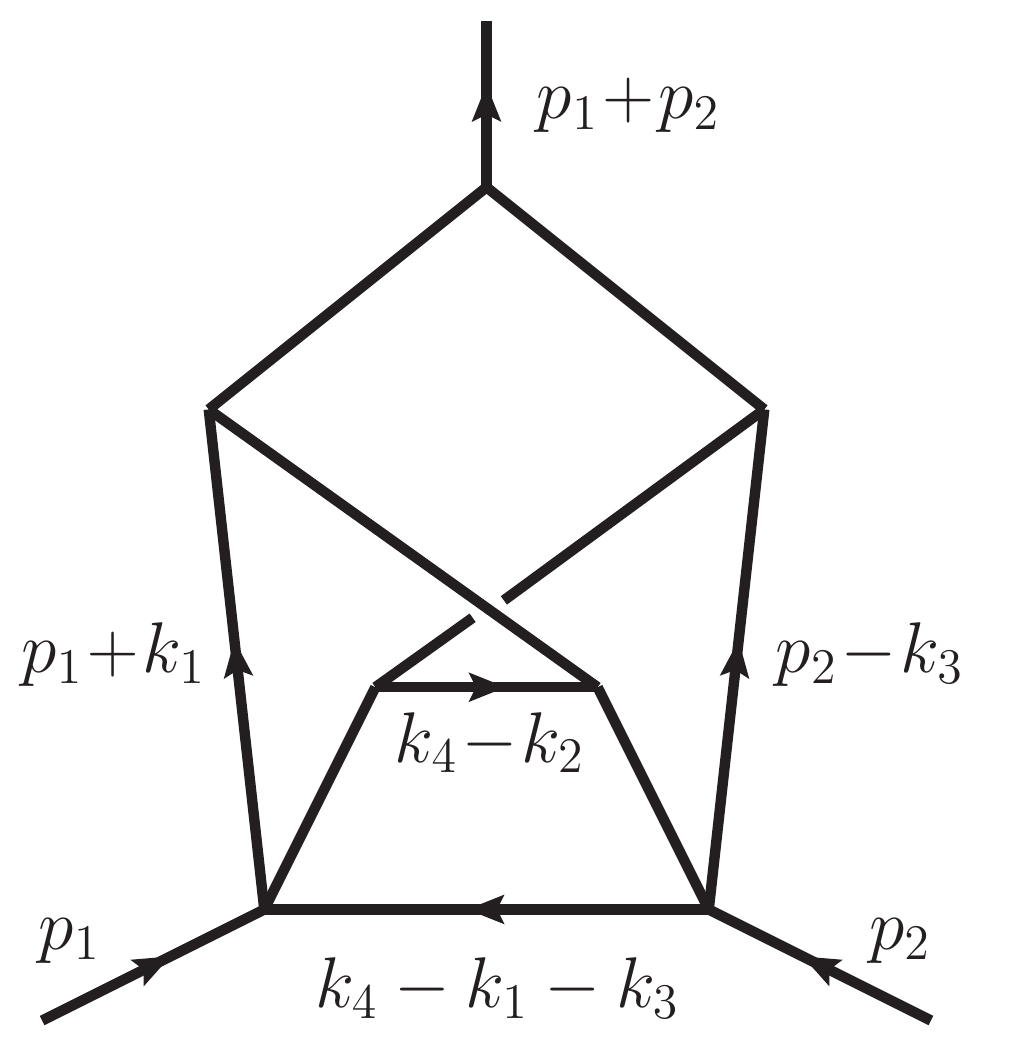}%
\caption{Typical form factor integral, which we compute for $p_1^2 \neq 0$. We then extract its value at $p_1^2=0$. }
\label{fig:exampleDE}
\end{figure}

For example, we find that the integral shown in Fig.~\ref{fig:exampleDE} for $p_1^2=p_2^2=0$ is given by
\begin{align}
\label{eq:Ic16}
I_{{\rm Fig.}\ref{fig:exampleDE}} &= 
\frac{1}{\epsilon ^2} \left[\frac{\zeta _3^2}{4}+\frac{31 \pi ^6}{30240}\right]
+ \frac{1}{\epsilon} \left[\frac{7 \pi ^4 \zeta _3}{180}-\frac{13 \pi ^2 \zeta _5}{16}-\frac{199 \zeta _7}{64}\right] \nonumber \\
&\hspace{-0.7cm}+\left[39 \zeta _{2,6}+\frac{13}{48} \pi ^2 \zeta _3^2+17 \zeta _3 \zeta _5-\frac{39301 \pi ^8}{1451520}\right] +\mathcal{O}(\eps)\,.
\end{align}
This agrees with our infrared analysis.

\section{Main results}

For the $n_f$ results quoted below we take the fermions to live in the fundamental representation ($X=F$), 
while for the $n_s$ and ${\cal N}=4$ sYM results we put all fields in the adjoint representation ($R=X=A$).
We find, 
\begin{align}
K_4|_\mathrm{quartic}^{n_f} &= n_f \CfourF
\bigg(\zeta_2 -\frac{\zeta_3}{3}-\frac{5 \zeta_5}{3} \bigg)
\label{eq:K4quarticQCD}
\,.
\end{align}
and
\begin{align}
K_4|_\mathrm{quartic}^{n_s} &= n_s \CfourA
\bigg(-\frac{\zeta_2}{4} +\frac{\zeta_3}{12} -\frac{5 \zeta_5}{24}
\bigg)
\label{eq:K4quarticQCDns}
\,.
\end{align}
The $n_f$ term is in perfect agreement with the numerical results of
\cite{Moch:2017uml,Moch:2018wjh}.

We can now combine our novel analytic result in \eq{K4quarticQCD},
 together with the full planar $n_f$ contribution~\cite{Lee:2016ixa},  the $n_f T_F C_R C_F^2$ term~\cite{Grozin:2018vdn},
 and a conjectured new result for the $n_f T_F C_R C_F C_A$ term from a parallel paper \cite{Bruser:2019auj},
to obtain the complete analytic (linear) $n_f$ term of the light-like QCD cusp anomalous dimension:
\begin{align}
K_4|^{n_f} =&\;
 n_f \CfourF   \bigg(\frac{\pi^2}{6} -\frac{\zeta_3}{3}-\frac{5 \zeta_5}{3} \bigg)
\nonumber\\
&+ n_f T_F C_R C_A^2   \bigg(-\frac{361 \zeta_3}{54}+\frac{7 \pi ^2
    \zeta_3}{36}+\frac{131 \zeta_5}{72} \nonumber\\
&\qquad -\frac{24137}{10368} +\frac{635 \pi ^2}{1944}-\frac{11 \pi ^4}{2160}\bigg)
\nonumber\\
&+ n_f T_F C_R C_A C_F  
\bigg(\frac{29 \zeta_3}{9}-\frac{\pi ^2 \zeta_3}{6}+\frac{5 \zeta_5}{4}-\frac{17033}{5184}
\nonumber\\
&\qquad +\frac{55 \pi ^2}{288}-\frac{11 \pi ^4}{720}\bigg)
\nonumber\\
&+n_f T_F C_R C_F^2  \bigg(\frac{37 \zeta_3}{24}-\frac{5 \zeta_5}{2}+\frac{143}{288}\bigg)
\,.
\end{align}
Note that all other fermionic contributions ($n_f^2$, $n_f^3$) are known~\cite{Beneke:1995pq,Grozin:2015kna,Davies:2016jie}.
Next, we use the numerical result of \cite{Moch:2018wjh} for the purely gluonic quartic
Casimir term, together with the analytic matter contributions computed here, to
obtain the result in ${\cal N}=4$ sYM,
\begin{align}
 K_4|_\mathrm{quartic}^\mathrm{N=4SYM}
 &= (-6.11047 \pm 0.0078) \,\CfourA
\,.
\end{align}
This agrees perfectly with the result of \cite{Boels:2017ftb}, 
\begin{align}
 K_4|_\mathrm{quartic, \ Boels \ et\  al.}^\mathrm{N=4SYM}
 &=(-6.4 \pm 0.76)\,\CfourA
\,,
\end{align}
and improves the numerical precision by two decimal places.

\section{Discussion and Outlook}
In this Letter, we computed the matter-dependent contributions to the
quartic Casimir term of the four-loop light-like cusp anomalous dimension in QCD.  
Combining this with other results, we obtained the full four-loop 
$n_{f}$ dependence of the cusp anomalous dimension in QCD.
We also obtained a more precise numerical result for the cusp anomalous
dimension in $\mathcal{N}=4$ sYM.

Our calculation was considerably simplified by using a basis
of master integrals of uniform transcendental weight with improved soft and collinear properties.
In this way, we did not require any master integrals with 11 or more propagators.
When extending this method to more general integrals, other regions than soft and collinear ones may be relevant as well, such as e.g. Glauber regions.  
Note that in principle, it is possible to verify the predictions of this analysis analytically by sector decomposition~\cite{Binoth:2000ps},  
without having to perform the
numerical integration steps of the implementations~\cite{Borowka:2017idc,Smirnov:2015mct}.
Finally, we expect that the integrals constructed in this way may also be more stable numerically \cite{Boels:2017ftb,Boels:2017skl}.

\section*{Acknowledgments}
We wish to thank R. Br\"user, J.\ Hoff and B.\ Mistlberger for fruitful discussions.
This research received funding from the European Union's Horizon 2020 research and innovation programme 
under European Research Council grant agreement No 725110, {\it Novel
structures in scattering amplitudes}, and under the Marie Skłodowska-Curie grant agreement 746223.
It was also supported in part by the PRISMA cluster of excellence at JGU Mainz.
J.\ M.\ H. and T.\ P.  also wish to thank the Galileo Galilei Institute for hospitality during the workshop ``Amplitudes in the LHC era".
The authors gratefully acknowledge computing support of the HPC group at JGU Mainz.

Note added: while these results were being prepared for publication, the preprint \cite{Lee:2019zop} appeared. 
$K_4|_\mathrm{quartic}^{n_f}$, which is computed there from the form factor of a fermion current, agrees with our result.

\bibliographystyle{h-physrev} 

\bibliography{refscusp.bib}

\begin{thebibliography}{10}

\bibitem{Mueller:1981sg}
A.~H. Mueller,
\newblock Phys. Rept. {\bf 73}, 237 (1981).

\bibitem{Korchemsky:1991zp}
G.~P. Korchemsky and A.~V. Radyushkin,
\newblock Phys. Lett. {\bf B279}, 359 (1992), hep-ph/9203222.

\bibitem{Korchemskaya:1994qp}
I.~A. Korchemskaya and G.~P. Korchemsky,
\newblock Nucl. Phys. {\bf B437}, 127 (1995), hep-ph/9409446.

\bibitem{Korchemsky:1992xv}
G.~P. Korchemsky and G.~Marchesini,
\newblock Nucl. Phys. {\bf B406}, 225 (1993), hep-ph/9210281.

\bibitem{Sterman:1986aj}
G.~F. Sterman,
\newblock Nucl. Phys. {\bf B281}, 310 (1987).

\bibitem{Becher:2006mr}
T.~Becher, M.~Neubert, and B.~D. Pecjak,
\newblock JHEP {\bf 01}, 076 (2007), hep-ph/0607228.

\bibitem{Becher:2008cf}
T.~Becher and M.~D. Schwartz,
\newblock JHEP {\bf 07}, 034 (2008), 0803.0342.

\bibitem{Abbate:2010xh}
R.~Abbate, M.~Fickinger, A.~H. Hoang, V.~Mateu, and I.~W. Stewart,
\newblock Phys. Rev. {\bf D83}, 074021 (2011), 1006.3080.

\bibitem{Hoang:2015hka}
A.~H. Hoang, D.~W. Kolodrubetz, V.~Mateu, and I.~W. Stewart,
\newblock Phys. Rev. {\bf D91}, 094018 (2015), 1501.04111.

\bibitem{Stewart:2013faa}
I.~W. Stewart, F.~J. Tackmann, J.~R. Walsh, and S.~Zuberi,
\newblock Phys. Rev. {\bf D89}, 054001 (2014), 1307.1808.

\bibitem{Becher:2013xia}
T.~Becher, M.~Neubert, and L.~Rothen,
\newblock JHEP {\bf 10}, 125 (2013), 1307.0025.

\bibitem{Chen:2018pzu}
X.~Chen {\em et~al.},
\newblock Phys. Lett. {\bf B788}, 425 (2019), 1805.00736.

\bibitem{Beisert:2006ez}
N.~Beisert, B.~Eden, and M.~Staudacher,
\newblock J. Stat. Mech. {\bf 0701}, P01021 (2007), hep-th/0610251.

\bibitem{Henn:2016men}
J.~M. Henn, A.~V. Smirnov, V.~A. Smirnov, and M.~Steinhauser,
\newblock JHEP {\bf 05}, 066 (2016), 1604.03126.

\bibitem{Lee:2016ixa}
J.~Henn, R.~N. Lee, A.~V. Smirnov, V.~A. Smirnov, and M.~Steinhauser,
\newblock JHEP {\bf 03}, 139 (2017), 1612.04389.

\bibitem{Davies:2016jie}
J.~Davies, A.~Vogt, B.~Ruijl, T.~Ueda, and J.~A.~M. Vermaseren,
\newblock Nucl. Phys. {\bf B915}, 335 (2017), 1610.07477.

\bibitem{vonManteuffel:2016xki}
A.~von Manteuffel and R.~M. Schabinger,
\newblock Phys. Rev. {\bf D95}, 034030 (2017), 1611.00795.

\bibitem{Lee:2017mip}
R.~N. Lee, A.~V. Smirnov, V.~A. Smirnov, and M.~Steinhauser,
\newblock Phys. Rev. {\bf D96}, 014008 (2017), 1705.06862.

\bibitem{Moch:2017uml}
S.~Moch, B.~Ruijl, T.~Ueda, J.~A.~M. Vermaseren, and A.~Vogt,
\newblock JHEP {\bf 10}, 041 (2017), 1707.08315.

\bibitem{Boels:2017ftb}
R.~H. Boels, T.~Huber, and G.~Yang,
\newblock JHEP {\bf 01}, 153 (2018), 1711.08449.

\bibitem{Boels:2017skl}
R.~H. Boels, T.~Huber, and G.~Yang,
\newblock Phys. Rev. Lett. {\bf 119}, 201601 (2017), 1705.03444.

\bibitem{Moch:2018wjh}
S.~Moch, B.~Ruijl, T.~Ueda, J.~A.~M. Vermaseren, and A.~Vogt,
\newblock Phys. Lett. {\bf B782}, 627 (2018), 1805.09638.

\bibitem{ArkaniHamed:2010gh}
N.~Arkani-Hamed, J.~L. Bourjaily, F.~Cachazo, and J.~Trnka,
\newblock JHEP {\bf 06}, 125 (2012), 1012.6032.

\bibitem{Henn:2013pwa}
J.~M. Henn,
\newblock Phys. Rev. Lett. {\bf 110}, 251601 (2013), 1304.1806.

\bibitem{WasserMSc}
P.~Wasser,
\newblock M.Sc.  (2016),
  https://publications.ub.uni-mainz.de/theses/frontdoor.php?source
  opus=100001967.

\bibitem{Gehrmann:2011xn}
T.~Gehrmann, J.~M. Henn, and T.~Huber,
\newblock JHEP {\bf 03}, 101 (2012), 1112.4524.

\bibitem{Drummond:2010mb}
J.~M. Drummond and J.~M. Henn,
\newblock JHEP {\bf 05}, 105 (2011), 1008.2965.

\bibitem{Bourjaily:2011hi}
J.~L. Bourjaily, A.~DiRe, A.~Shaikh, M.~Spradlin, and A.~Volovich,
\newblock JHEP {\bf 03}, 032 (2012), 1112.6432.

\bibitem{Kotikov:1990kg}
A.~V. Kotikov,
\newblock Phys. Lett. {\bf B254}, 158 (1991).

\bibitem{Bern:1993kr}
Z.~Bern, L.~J. Dixon, and D.~A. Kosower,
\newblock Nucl. Phys. {\bf B412}, 751 (1994), hep-ph/9306240.

\bibitem{Gehrmann:1999as}
T.~Gehrmann and E.~Remiddi,
\newblock Nucl. Phys. {\bf B580}, 485 (2000), hep-ph/9912329.

\bibitem{Henn:2013nsa}
J.~M. Henn, A.~V. Smirnov, and V.~A. Smirnov,
\newblock JHEP {\bf 03}, 088 (2014), 1312.2588.

\bibitem{vanRitbergen:1997va}
T.~van Ritbergen, J.~A.~M. Vermaseren, and S.~A. Larin,
\newblock Phys. Lett. {\bf B400}, 379 (1997), hep-ph/9701390.

\bibitem{Chetyrkin:1981qh}
K.~G. Chetyrkin and F.~V. Tkachov,
\newblock Nucl. Phys. {\bf B192}, 159 (1981).

\bibitem{vonManteuffel:2014ixa}
A.~von Manteuffel and R.~M. Schabinger,
\newblock Phys. Lett. {\bf B744}, 101 (2015), 1406.4513.

\bibitem{Peraro:2016wsq}
T.~Peraro,
\newblock JHEP {\bf 12}, 030 (2016), 1608.01902.

\bibitem{Lee:2012cn}
R.~N. Lee,
\newblock (2012), 1212.2685.

\bibitem{Hoff:2016pot}
J.~Hoff,
\newblock J. Phys. Conf. Ser. {\bf 762}, 012061 (2016), 1607.04465.

\bibitem{Arkani-Hamed:2014via}
N.~Arkani-Hamed, J.~L. Bourjaily, F.~Cachazo, and J.~Trnka,
\newblock Phys. Rev. Lett. {\bf 113}, 261603 (2014), 1410.0354.

\bibitem{Chicherin:2018old}
D.~Chicherin {\em et~al.},
\newblock (2018), 1812.11160.

\bibitem{Eden:2012tu}
B.~Eden, P.~Heslop, G.~P. Korchemsky, and E.~Sokatchev,
\newblock Nucl. Phys. {\bf B862}, 450 (2012), 1201.5329.

\bibitem{Anastasiou:2018rib}
C.~Anastasiou and G.~Sterman,
\newblock (2018), 1812.03753.

\bibitem{Lee:2011jt}
R.~N. Lee, A.~V. Smirnov, and V.~A. Smirnov,
\newblock Nucl. Phys. {\bf B856}, 95 (2012), 1108.0732.

\bibitem{Smirnov:2008iw}
A.~V. Smirnov,
\newblock JHEP {\bf 10}, 107 (2008), 0807.3243.

\bibitem{Remiddi:1999ew}
E.~Remiddi and J.~A.~M. Vermaseren,
\newblock Int. J. Mod. Phys. {\bf A15}, 725 (2000), hep-ph/9905237.

\bibitem{Maitre:2005uu}
D.~Maitre,
\newblock Comput. Phys. Commun. {\bf 174}, 222 (2006), hep-ph/0507152.

\bibitem{Grozin:2018vdn}
A.~Grozin,
\newblock JHEP {\bf 06}, 073 (2018), 1805.05050.

\bibitem{Bruser:2019auj}
R.~Brüser, A.~Grozin, J.~M. Henn, and M.~Stahlhofen,
\newblock JHEP {\bf 05}, 186 (2019), 1902.05076.

\bibitem{Beneke:1995pq}
M.~Beneke and V.~M. Braun,
\newblock Nucl. Phys. {\bf B454}, 253 (1995), hep-ph/9506452.

\bibitem{Grozin:2015kna}
A.~Grozin, J.~M. Henn, G.~P. Korchemsky, and P.~Marquard,
\newblock JHEP {\bf 01}, 140 (2016), 1510.07803.

\bibitem{Binoth:2000ps}
T.~Binoth and G.~Heinrich,
\newblock Nucl. Phys. {\bf B585}, 741 (2000), hep-ph/0004013.

\bibitem{Borowka:2017idc}
S.~Borowka {\em et~al.},
\newblock Comput. Phys. Commun. {\bf 222}, 313 (2018), 1703.09692.

\bibitem{Smirnov:2015mct}
A.~V. Smirnov,
\newblock Comput. Phys. Commun. {\bf 204}, 189 (2016), 1511.03614.

\bibitem{Lee:2019zop}
R.~N. Lee, A.~V. Smirnov, V.~A. Smirnov, and M.~Steinhauser,
\newblock JHEP {\bf 02}, 172 (2019), 1901.02898.

\end{thebibliography}
\end{document}